\documentclass[letterpaper,english,reprint,showpacs,aps,nofootinbib,prl]{revtex4-1}
\usepackage[utf8]{inputenc}
\usepackage{amsmath}
\usepackage{amssymb}
\usepackage{graphicx}



\usepackage{babel}
\begin{document}

\title{Microwave excitation of spin wave beams in thin ferromagnetic films}

\author{P. Gruszecki}
\email{pawel.gruszecki@amu.edu.pl}
\selectlanguage{english}%
\affiliation{Faculty of Physics, Adam Mickiewicz University in Pozna\'n, Umultowska
85, Pozna\'n, 61-614, Poland }

\author{A.~E.~Serebryannikov}
\affiliation{Faculty of Physics, Adam Mickiewicz University in Pozna\'n, Umultowska
85, Pozna\'n, 61-614, Poland }

\author{M. Krawczyk}
\email{krawczyk@amu.edu.pl}

\selectlanguage{english}%
\affiliation{Faculty of Physics, Adam Mickiewicz University in Pozna\'n, Umultowska
85, Pozna\'n, 61-614, Poland }

\author{W. \'Smigaj}
\affiliation{Simpleware Ltd., Bradninch Hall, Castle Street, Exeter, EX4 3PL,
UK}

\begin{abstract}
We present an approach enabling generation of narrow spin wave beams
in thin homogeneous ferromagnetic films. The main idea is to match
the wave vector of the spin wave with that corresponding to the spectral
maximum of the exciting microwave magnetic field only \emph{locally},
in the region of space from which the beam should be launched. We
show that this can be achieved with the aid of a properly designed
coplanar waveguide transducer generating a nonuniform microwave magnetic
field. The resulting two-dimensional spin wave beams obtained in micromagnetic
simulations propagate over distances of several micrometers. The proposed
approach requires neither inhomogeneity of the ferromagnetic film
nor nonuniformity of the biasing magnetic field, and it can be generalized
to yield multiple spin wave beams of different width at the same frequency.
Other possible excitation scenarios and applications of spin wave
beam magnonics are also discussed. 
\end{abstract}
\maketitle
Magnonics is an emerging field of research and technology \citep{Kruglyak10}
whose area of interest is spin wave (SW) dynamics. It is closely related
to photonics, which deals with electromagnetic waves, and phononics,
which is concerned with elastic waves. Generation \citep{Demidov09,Au13}, transmission \citep{Kajiwara10,Vogt14}, signal processing \citep{Klingler14}, amplification \citep{Bracher14}, and detection \citep{Vlaminck08} of SWs are major subjects of study in magnonics. Basic components performing these operations in nano- and mesoscale have been recently demonstrated. However, their functionality and performance require significant improvement to make magnonic elements competitive (in terms of energy efficiency, throughput, etc.) with other kinds of integrated devices: electronic, photonic or acoustic \citep{Bernstein10,Nikonov13}. The theoretical underpinnings of magnonics also constitute a rich and still not fully explored research area. SW caloritronics \citep{An13}, magnonics-spintronics \citep{Madami11}, magnonic crystals \citep{Krawczyk14}, magnonic metamaterials \citep{Mruczkiewicz12}, and Bose-Einstein
condensates of magnons \citep{Demidov08} are some of the topics whose
investigations have been started in recent years.

The crucial milestone in modern photonics (both from the scientific
and the technological perspective) was the development of efficient
sources of coherent light beams: lasers. In magnonics, an efficient
source of coherent SW beams is still not available, in spite of many
attempts at its development. The caustic effect \citep{Schneider10},
nonlinear self-focusing of SWs \citep{Boyle96,Bauer97}, and inhomogeneous internal magnetic field (the sum of all magnetic fields inside magnetic material) \citep{Bini75} are among the beam excitation mechanisms investigated to date. The first is limited to low frequencies, i.e.,
to magnetostatic SWs, which have a caustic dispersion relation. The
excitation is limited to specific angles with respect to the direction
of the magnetization vector, which needs to be in the plane of the
film. The second approach requires excitation of SWs with high amplitude
and the beam spreads quickly after passing the focal point. In the
third approach, it is necessary to create an inhomogeneity of the
internal magnetic field, i.e., to enclose SW excitations to propagate within channels
of the static magnetic field with decreased magnitude, thus technologically complex.

On the other hand, electromagnetic wave transducers---microstripe
and coplanar waveguides (CPWs)---are extensively used in studies of
SWs in thin ferromagnetic films \citep{Vlaminck10,Maksymov15}, in
particular as sources of plane-wave-like SWs. In this paper, we propose
and numerically validate with micromagnetic simulations (MSs) a method
of excitation of narrow SW beams in homogeneous ferromagnetic thin
films using the magnetic field generated by microwave-frequency transducers
with a suitable geometry. We demonstrate that efficient excitation
of a SW beam can occur only within those space regions where the wave
vectors of the SW and the Fourier-transformed distribution of the
microwave (mf) magnetic field match. To ensure that this condition
holds only locally, we vary the transducer's profile along the current
flow direction. This affects the spatial distribution of the magnetic
field induced by the current and, in particular, its Fourier spectrum
along the expected direction of SW propagation. In its most basic
form, this method enables excitation of a single well-localised SW
beam propagating perpendicularly to the waveguide axis. By adapting
the transducer's geometry, this approach can be generalised so that
multiple SW beams of identical or different width appear simultaneously
in different regions of the ferromagnetic film. Other possible scenarios
and applications are also discussed.

Throughout this paper, we consider the propagation of SWs in a ferromagnetic
film of thickness $t_{\text{f}}=20$~nm made of yttrium-iron garnet
(YIG). YIG is a promising dielectric material with the lowest damping
of SWs ever recorded. Recently, the technology of ultra-thin YIG film
deposition has been developed, holding promise for integrated magnonics
\citep{Sun12,Yu14}. We assume the film is saturated normal to its
plane, i.e., along the $z$~axis, by an external magnetic field $\mu_{0}H=1$~T,
where $\mu_{0}$ is the permeability of vacuum. We take the YIG magnetization
saturation to be $M_{\text{S}}=0.194\times10^{6}$~A/m, exchange
constant $A=0.4\times10^{-11}$~J/m, and gyromagnetic ratio $\gamma=176$~rad~GHz/T
\citep{Stancil}.

The results of MSs reported in this paper have been obtained with
the open-source software MuMax3 employing the finite difference method
\citep{mumax}. We used a finite difference grid with resolution 5~nm
in the $x$ and~$y$ directions (i.e., in the film plane) and 20~nm
in the $z$ direction (along the normal to the film surface) and the
Landau-Lifshitz equation was solved using the RK45 (Dormand-Prince)
method\citep{Dormand80,comment1}. The distributions of the magnetic
fields generated by CPWs and used to excite the SWs have been calculated
with CST Microwave Studio, a commercial Maxwell solver based on the
finite integration method \citep{CST}.

\begin{figure}[!ht]
\includegraphics[width=8.5cm]{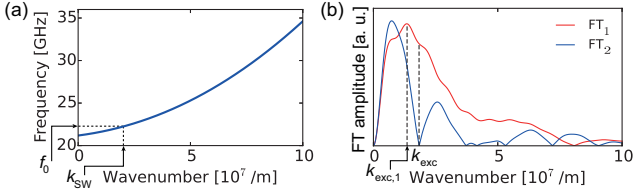} \protect\protect\protect\protect\caption{(a) The dispersion relation of SWs in a 20-nm thick YIG film with
out-of-plane magnetization placed in an external magnetic field of
1 T. (b) The Fourier transforms FT$_{1}$ and FT$_{2}$ of the microwave
magnetic fields $h_{y,\text{1}}^{\mathrm{mf}}(y)$ and $h_{y,\text{2}}^{\mathrm{mf}}(y)$
emitted by two CPWs, CPW$_{1}$ and CPW$_{2}$, at frequency $f_{0}$.
$k_{\text{exc,1}}$ is the location of the maximum of $\mbox{FT}_{\text{1}}$.
At $k_{\text{exc}}=k_{\text{SW}}(f_{0})$, the wave number of SWs
at frequency $f_{0}$, the value of $\mbox{FT}_{\text{1}}$ is still
close to maximal, while $\mbox{FT}_{\text{2}}$ has a minimum. \label{fig:F1} }
\end{figure}

Let us now discuss the basic principle of the selective (local) wave
vector matching. Following \citep{Vlaminck08}, we assume that the
efficiency of monochromatic SW excitation by a CPW depends, in particular,
on the quality of the match between the wave number $k_{\text{SW}}(f_{0})$
of the SW at the excitation frequency $f=f_{0}$ and the wave number
$k_{\mathrm{exc}}$ corresponding to the maximum of the Fourier transform
(FT) of the microwave magnetic field $\mathbf{h}^{\mathrm{mf}}$ oscillating
at the same frequency. Thus, the most efficient excitation of SWs
occurs when 
\begin{equation}
k_{\mathrm{SW}}(f_{0})=k_{\mathrm{exc}}.\label{Eq:BasicCond}
\end{equation}

On the other hand, if the magnitude of the FT at a wave number $k=k_{\mathrm{nexc}}$
is very small, there is no matching and no efficient SW excitation
should arise. These facts make it possible to engineer the wavefronts
of excited SWs, and in particular to excite SW beams with high efficiency.

Figure~\ref{fig:F1}(a) shows the dispersion relation of SWs in a
YIG film of thickness $t_{\text{f}}=20$~nm, calculated with the
analytical formula \citep{Stancil} 
\begin{equation}
\begin{split}f(k_{\mathrm{SW}}) & =\frac{\gamma\mu_{0}}{2\pi}\left[\left(H-M_{\text{S}}+\frac{2A}{\mu_{0}M_{\text{S}}}k_{\mathrm{SW}}^{2}\right)\right.\\
 & \quad\times\left.\left(H+\frac{2A}{\mu_{0}M_{\text{S}}}k_{\text{SW}}^{2}-M_{\text{S}}\frac{1-\text{e}^{-k_{\text{SW}}t_{\text{f}}}}{k_{\text{SW}}t_{\text{f}}}\right)\right]^{\frac{1}{2}}.
\end{split}
\label{Eq:Dispersion}
\end{equation}
In line with Eq.~(\ref{Eq:Dispersion}), the dispersion is independent
from the direction of the wave vector and has a predominantly parabolic
behavior in the considered wave number range.

For reasons discussed below, SWs are excited mainly by the microwave
magnetic field component parallel to the expected direction of their
propagation, further denoted by $y$. Thus, we focus on the behavior
of $\mathcal{F}\left[h_{y}^{\mathrm{mf}}(y)\right]$, the FT of $h_{y}^{\mathrm{mf}}(y)$.
Fig.~\ref{fig:F1}(b) shows the Fourier transforms $\mathrm{FT_{1}}=\mathcal{F}\left[h_{y,\text{1}}^{\mathrm{mf}}(y)\right]$
and $\mathrm{FT}_{2}=\mathcal{F}\left[h_{y,\text{2}}^{\mathrm{mf}}(y)\right]$
of the magnetic fields generated by two realistic uniform transducers.
Their geometry will be detailed later in the paper. At frequency $f_{0}=22.12$~GHz
\citep{comment2}, Eq.~(\ref{Eq:BasicCond}) holds only for FT$_{1}$
{[}$k_{\text{exc}}=1.8\times10^{7}$ m$^{-1}$$\approx k_{\text{SW}}(f_{0})$,
see Fig.~\ref{fig:F1}{]}; therefore one may expect that at this
frequency SWs can be excited only with the field $h_{y,\text{1}}^{\mathrm{mf}}(y)$.
the transducer is designed so that it produces the magnetic field
distribution $h_{y,2}^{\mathrm{mf}}(y)$ over most of its length and
$h_{y,1}^{\mathrm{mf}}(y)$ only along a short section, an SW will
be excited only in that short section of the transducer. Provided
that this section is at least a few times longer than the SW wavelength,
this should give rise to a well-collimated SW beam. Beam excitation
in a wide frequency band will be possible if the dip in FT$_{2}$
near $k_{\text{exc}}=k_{\text{SW}}$ is sufficiently broad and the
slope of the SW dispersion relation near $k_{\text{exc}}$ sufficiently
steep.

\begin{figure}[!ht]
\includegraphics[width=8cm]{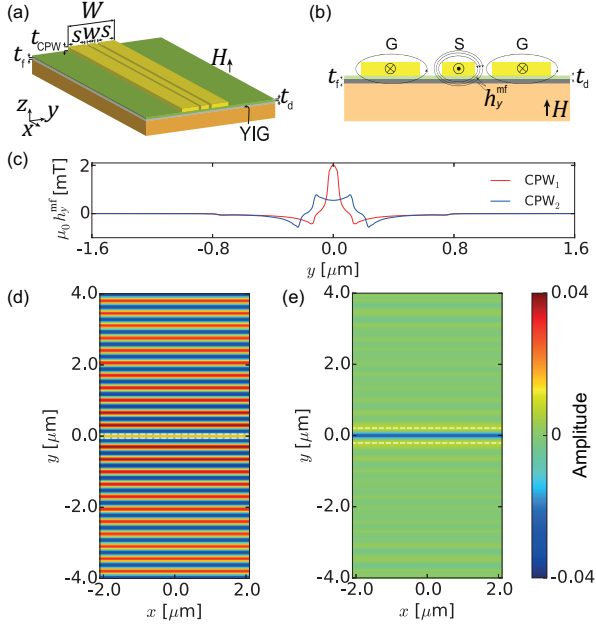}

\protect\protect\caption{(a)~Perspective view and (b)~cross-section of a CPW consisting of
a signal line (denoted by S) and two ground lines (denoted by G).
A ferromagnetic (YIG) film of thickness $t_{\text{f}}$ is separated
from the CPW by a nonmagnetic, dielectric layer of thickness $t_{\text{d}}$.
The YIG is saturated normal to the film plane by an external magnetic
field $H$. The microwave signal transmitted along the $x$~axis
generates a magnetic field $\mathbf{h^{\mathrm{mf}}}$, which exerts
a torque on the magnetization in YIG. (c)~The $y$~component of
the microwave magnetic field along the $y$ axis, $h_{y}^{\mathrm{mf}}(y)$,
excited by $\text{CPW}_{1}$ (red line) and $\text{CPW}_{2}$ (blue
line), obtained from CST simulations. (d)--(e) Dynamic component of
the magnetization vector of the SWs induced by the microwave magnetic
field from (d) $\text{CPW}_{1}$ and (e) $\text{CPW}_{2}$, obtained
from MSs. The horizontal white dashed lines correspond to the signal
line width. \label{fig:F2}}
\end{figure}

Now, let us briefly revisit SW excitation by the microwave magnetic
field induced by a current flowing through the uniform transducer
shown in Fig.~\ref{fig:F2}(a) \citep{Maksymov15}. The CPW consists
of a signal line of width~$w$ separated by gaps of width $s$ from
two identical ground lines. All lines are deposited on a nonmagnetic,
dielectric layer of thickness $t_{\mathrm{d}}$ covering the ferromagnetic
film. A microwave signal in the CPW generates the field $\mathbf{h}^{\mathrm{mf}}$,
which oscillates in the $(y,z)$ plane, i.e., perpendicular to the
CPW axis, see Fig.~\ref{fig:F2}(b). In the plane of the ferromagnetic
film, only the $y$~component of this field exerts a non-zero torque
on the magnetization, which is parallel to the $z$~axis. This torque
induces coherent magnetization precession around the equilibrium direction
and causes generation of SWs propagating along the $y$~axis.

Figure \ref{fig:F2}(c) shows the profile of the microwave magnetic
field 7.5~nm below the ground and signal lines of two CPWs made of
Cu with conductivity $\sigma=5.88\times10^{7}$ S/m \citep{Kittel},
nm, operating at $f_{0}=22.12$~GHz. The dielectric layer has thickness
$t_{\text{d}}=5$~nm. The width of the signal line in $\text{CPW}_{1}$
is $w_{1}=65$~nm and in $\text{CPW}_{2}$, $w_{2}=260$~nm. The
total width of the transducers and the width of the gap between the
signal and ground lines are the same in both waveguides: $W_{1}=W_{2}=1.5$~$\mu$m
and $s_{1}=s_{2}=85$~nm. By Fourier-transforming the field distributions
from Fig.~\ref{fig:F2}(c) we obtain the curves FT$_{1}$ and FT$_{2}$
from Fig.~\ref{fig:F1}(b). This shows that the desired behavior
of $h_{y}^{\mathrm{mf}}(y)$ in the ferromagnetic film can be achieved
by taking advantage of the sensitivity of the field profile of the
CPW mode to the signal line width~$w$. Indeed, Figs.\ 2(d) and~2(e)
show the SWs induced by $\text{CPW}_{1}$ and $\text{CPW}_{2}$ at
$f_{0}=22.12$ GHz, i.e., when $k_{\text{SW}}(f_{0})\approx k_{\mathrm{exc}}$
for the former and $k_{\text{SW}}(f_{0})\neq{k_{\mathrm{exc}}}$ ($k_{\text{SW}}(f_{0})\approx k_{\mathrm{nexc}}$)
for the latter. Note that in CPW$_{2}$ $\mathcal{F}\left[h_{y,\text{2}}^{\mathrm{mf}}(y)\right]$
is close to 0 at $k=k_{\text{SW}}(f_{0})$, see Fig.~\ref{fig:F1}(b).
As expected, propagating SWs are excited only in CPW$_{1}$, in which
Eq.~(\ref{Eq:BasicCond}) is satisfied.

\begin{figure}[!ht]
\includegraphics[width=8cm]{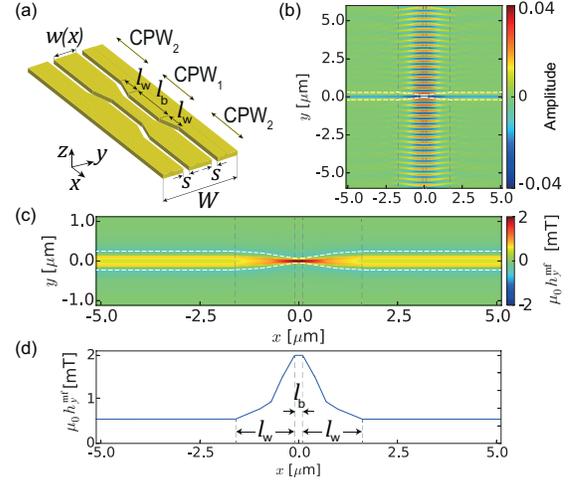} \protect\protect\caption{(a) Geometry of the nonuniform CPW proposed to excite an SW beam.
(b) Dynamic component of the magnetization vector of the SW beam excited
in a thin YIG film by the microwave magnetic field {[}shown in plot
(c){]} induced by the nonuniform CPW (results of MSs). (c) Distribution
of the microwave magnetic field $h_{y}^{\mathrm{mf}}$ induced by
the CPW on the top surface of the YIG film. The shape of the signal
line is marked with dashed white lines. (d) Profile of $h_{y}^{\mathrm{mf}}$
along the axis of the signal line, $y=0$.\label{fig:F3}}
\end{figure}

To restrict SW generation to a short section of the transducer and
hence to obtain a clearly defined beam, we vary the CPW geometry along
the $x$~axis. The signal line width is changed from $w_{2}$ at
$x=-(l_{\text{w}}+l_{\text{b}}/2)$ to $w_{1}$ at $x=-l_{\text{b}}/2$
and back from $w_{1}$ at $x=l_{\text{b}}/2$ to $w_{2}$ at $x=l_{\text{w}}+l_{\text{b}}/2$,
as shown in Fig.~\ref{fig:F3}(a). The gap width $s=190$~nm and
the total CPW width $W=3$~$\mu$m are kept constant. In calculations,
we take $l_{\mathrm{w}}=1.5$~$\mu$m and $l_{\mathrm{b}}=200$~nm.
The microwave magnetic field distribution depends now also on the
$x$~coordinate: $h_{y}^{\text{mf}}=h_{y}^{\text{mf}}(x,y)$. To
match the mesh cells of MSs, the field used in MSs was constructed
by interpolating field profiles $h_{y}^{\text{mf}}(x_{j},y)$ obtained
from CST simulations at several values $x_{j}$ of $x$ (where $x_{j+1}-x_{j}=150$~nm),
using the following linear homotopic transformation providing a linear
change of the line width between points $x_{j}$ and $x_{j+1}$: 
\begin{equation}
\begin{split}h_{y}^{\text{mf}}(x,y) & =h_{y}^{\text{mf}}\left(x_{j},y\frac{y_{0,j}}{B(x,y)}\right)\\
 & \quad+c(x)\left[h_{y}^{\text{mf}}\left(x_{j+1},y\frac{y_{0,j+1}}{B(x,y)}\right)\right.\\
 & \quad-\left.h_{y}^{\text{mf}}\left(x_{j},y\frac{y_{0,j}}{B(x,y)}\right)\right],
\end{split}
\end{equation}
where $B(x,y)=y_{0,j}+c(x)(y_{0,j+1}-y_{0,j})$ and $c(x)=(x-x_{j})/(x_{j+1}-x_{j})$.
The symbol $y_{0,j}$ denotes the zero of $h_{y}^{\text{mf}}(x_{j},y)$
on the half-line $y>0$. This approach was introduced to conserve
the proper position of zeros of the field in the transition region.
The map of the $y$~component of the microwave magnetic field generated
by this nonuniform CPW on the top surface of the ferromagnetic film
is shown in Fig.~\ref{fig:F3}(c). 

Figure~\ref{fig:F3}(d) shows the distribution of the dynamic magnetic
field $h_{y}^{\mathrm{mf}}$ at the top surface of the YIG film along
the $x$ axis, i.e., the symmetry axis of the signal line. The dynamic component of the magnetization vector of a SW generated
at $22.12$~GHz with the aid of this CPW in the homogeneous YIG film
is plotted in Fig.~\ref{fig:F3}(b). The beam-type behavior of the
SW is evident, so that the basic idea of the suggested approach is
justified. As expected, the beam remains well convergent even at large
distances from the transducer. The SW beam width at the half-power
level is $1.5$~$\mu$m in the beam waist lying at $y=0$. This width
can be controlled by adjusting the length of the transducer section
having the profile of $\text{CPW}_{1}$, $l_{\mathrm{b}}$, and the
size and curvature of the transition section, $l_{\mathrm{w}}$, as
will be shown in the next paragraph. The minimum achievable beam width
is likely to be restricted by the available nanofabrication technology.

\begin{figure}[!t]
\includegraphics[width=8cm]{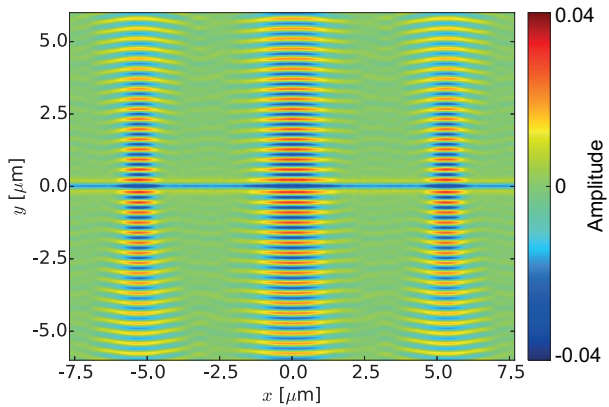} \protect\protect\protect\protect\caption{Dynamic component of the magnetization vector in three SW beams formed
simultaneously in a thin YIG film by the microwave magnetic field
generated by a single complex-shaped CPW (results of MSs). The left
and right beams, generated by a nonuniform CPW with $l_{\text{b}}=0$~nm
and $l_{\text{w}}=0.9$~$\mu$m, are narrower than the middle beam,
generated by a CPW with $l_{\text{b}}=100$~nm and $l_{\text{w}}=1.8$~$\mu$m.
\label{fig:F4} }
\end{figure}

Since the basic mechanism of SW beam generation in YIG films has now
been numerically validated, it is worthwhile to consider further excitation
scenarios. For example, multiple modulated sections can be introduced
into a CPW to generate multiple coherent SW beams in a homogeneous
ferromagnetic film. An example of SW excitation in a multibeam regime
is presented in Fig.~\ref{fig:F4}. As the beam divergence is weak,
individual beams do not interfere with each other in the considered
space region. Each beam may have different characteristics, such as
its width and intensity, since they are determined by the geometry
of the local CPW perturbation. This opens a route to multichannel
structures for SWs created simply by modulating the microwave magnetic
field through local modifications of the transducer geometry, with
no need to make either the ferromagnetic film or the biasing magnetic
field nonuniform.

The possibilities discussed above do not exhaust the broad range of
potential applications of SW beam magnonics. For instance, demultiplexer-type
operation might be achieved by creating such a distribution of $h_{y}^{\mathrm{mf}}(x,y)$
that injection of a mixed dual-frequency microwave signal into the
CPW would cause efficient generation of SW beams at two distinct frequencies
within different regions of the YIG film, associated with different
virtual propagation channels. However, to realize this regime, additional
design efforts are required, e.g., to limit excitation at undesirable
frequencies in both channels. 

SW beams can also be utilized for sensing, transmitting signals, or
performing logic operations. In particular, the concepts of logic
elements based on interference effects, such as Mach-Zehnder interferometers,
can also be adapted for SW-beam-based operation, provided that a mechanism
for introducing a controllable phase shift between two SW beams is
developed. Reading can be done by direct beam interference \citep{Hertel04}
(this will require a change of the beam propagation direction) or
using a detecting CPW transducer of appropriate width \citep{Schneider08,Sato13}.

To summarize, we have proposed to use nonuniform microwave CPW transducers
to excite effectively even high frequency SW beams in thin YIG films. Modulation of the transducer
geometry ensures that the necessary condition of SW generation, the
match between the SW wave vector and the maximum of the spatial Fourier
transform of the microwave magnetic field induced by the transducer,
is satisfied only locally, in a short section of the waveguide. The
beam characteristics, such as its waist width, are controlled by the
geometry of the modulated transducer section. The presented numerical
examples demonstrate the potential of the proposed approach, which
holds promise as a platform for future SW-based circuitry. In particular,
we have shown that multiple SW beams associated with separate virtual
propagation channels can be generated in a single thin ferromagnetic
film. Manipulation of such beams in particular magnonic devices will
be the subject of upcoming research. 
\begin{acknowledgments}
The research received funding from the National Science Center of
Poland UMO-2012/07/E/ST3/00538 and from the European Union Horizon
2020 research and innovation programme under the Marie Sklodowska-Curie
grant agreement No 644348. Part of the calculations presented in this
paper was performed at the Poznan Supercomputing and Networking Center.
AES thanks his colleagues from the Bilkent University, Ankara, for
help with CST simulations. PG would like to thank L.~Nizio for fruitful
discussions. \end{acknowledgments}


\begin{thebibliography}{99}
\bibitem{comment2} This frequency is achievable by standard microwave
generators, and at this frequency the dispersion relation of SWs in
YIG thin film is almost isotropic also when the magnetization saturation
is in-plane of the film. This makes the results of our paper more
general. However, at small enough wave vectors the strong anisotropy
of the dispersion relation has to be taken into account.
\end{thebibliography}
\end{document}